\documentclass[12pt]{revtex4-1}

\usepackage{amsmath,amssymb}
\usepackage{graphicx}
\usepackage{gensymb}
\usepackage[all]{xy}
\usepackage{epstopdf}

\newcommand{\ket}[1]{\mathop{\left|#1\right>}\nolimits}            
\newcommand{\braket}[2]{\langle #1 | #2 \rangle}
\newcommand{\ketbra}[2]{| #1\rangle\!\langle #2 |}

\begin{document}

\title{Lorentz transformations for massive two-particle systems: entanglement change and invariant subspaces}

\author{Esteban Castro-Ruiz and Eduardo Nahmad-Achar}

\address{Instituto de Ciencias Nucleares,
Universidad Nacional Aut\'onoma de M\'exico \\
Apdo. Postal 70-543, Mexico D.F., C.P. 04510}

\begin{abstract}
\noindent 
Lorentz boosts on particles with spin and momentum degrees of freedom induce momentum-dependent rotations. Since, in general, different particles have different momenta, the transformation on the whole state is not a representation of the rotation group. Here we identify the group that acts on a two-particle system and, for the case when the momenta of the particles are correlated, find invariant subspaces that have interesting properties for quantum information processes in relativistic scenarios. A basis of states is proposed for the study of transformations of spin states under Lorentz boosts, which is a good candidate for building quantum communication protocols in relativistic scenarios.

\vspace{0.3in}
{\it nahmad@nucleares.unam.mx}

\end{abstract}



\maketitle


\section{Introduction}
\label{one}
\noindent The conceptual basis of Quantum Mechanics has been the subject of heated debates since the beginning of the theory until these days. The EPR \cite{EPR} and Bohm \cite{Bohm} thought experiments set the context of the discussion on the nature of the physical reality, as understood according to Quantum Theory. Later, the works by Bell \cite{Bell} allowed for this discussion, previously philosophical, to be the subject to empirical verification. Experimental results arising from Bell's work \cite{Clauser, Aspect, GHZ} strongly suggest the impossibility for a local hidden variable theory to reproduce the predictions of Quantum Mechanics.

The study of the foundations of the theory has also led to the analysis of quantum systems as information carriers. The features that distinguish quantum systems from classical ones can be used to transmit and process information in ways that are impossible for systems that obey classical laws. Typical examples of this fact are quantum teleportation \cite{Teleportation} and super-dense coding \cite{SuperDenseCoding}. In order to characterise the novel properties of quantum mechanical systems in a precise manner, there has been an important amount of effort in defining  quantities that measure those aspects of the systems that are relevant to the transmission and processing of information (cf., e.g., \cite{Bengtsson, Holevo}). This analysis has been important both from the fundamental and the applied points of view.

Recently, there has been considerable interest in studying the behaviour of quantum systems under relativistic transformations from a quantum information perspective. Given the non-local character of quantum mechanics, experiments which produce non-local correlations have been analysed in a special-relativistic framework \cite{Czachor, Friis, Hacyan1, Hacyan2, PeresTernoRMP, SaldanhaVedralNJP, SaldanhaVedralPRA}. Furthermore, in view of the important applications of quantum systems for the transmission of information, the effects that relativistic velocities between emitter and receiver have on the capacity of quantum channels to transmit classical or quantum information have been studied \cite{BradlerCastro-RuizNahmad-Achar}, showing that an appropriate Lorentz boost increases both the classical and quantum capacities of a communication channel. Moreover, it can be used to obtain a positive quantum capacity channel from a channel that does not allow quantum communication for observers at rest with respect to each other.
 
In a similar spirit, the effects of Lorentz transformations on the quantum entanglement of systems of pairs of particles have been analysed \cite{Alsing, Friis, Castro-RuizNahmad-Achar}. Despite the fact that the total entanglement of the system, i.e. entanglement of one particle with respect to the other, is conserved, entanglement in the spin sector of a two-particle system is modified by the relative velocity between observers. In~\cite{Castro-RuizNahmad-Achar} it was shown that, for particles propagating with opposite momenta and a Lorentz boost in a perpendicular direction to the momenta, there exists a set of spin states that remain invariant under the Lorentz transformation, thereby conserving entanglement with respect to all partitions of the Hilbert space of the system. 

In this paper we consider the transformation of a two particle state in the general case and analyse closely the characteristics of invariant subspaces. 
In section~\ref{two} we briefly review how Lorentz transformations affect elementary particles with momentum and spin degrees of freedom and in section~\ref{three} we focus on two-particle systems and identify the group that acts on the spin sector of the state. We then analyse the case  where the momenta of the particles are correlated and show that there are subspaces of spin states that are closed under Lorentz transformations. In section~\ref{four} we give a closed formula for the spin-momentum entanglement of an EPR-like pair of arbitrary spin under a Lorentz boost perpendicular to the propagation direction. We present our conclusions in section~\ref{five}.

\section{One particle}
\label{two}
\noindent We briefly recall the transformation law for momentum and spin eigenstates under the Lorentz group. For a particle of mass $M>0$ and spin $s$, the state of momentum $p$ and spin projection along the $z$-axis $\sigma$ is defined as
\begin{equation}
\label{WignerBasis}
\ket{p, \sigma} =  U(L_p)\ket{k, \sigma},
\end{equation}
where $k = (M,0,0,0)^T$ is the four-momentum of the particle in its rest frame, $\sigma = -s, -s+1, \cdot \cdot\cdot, s$, and $U(L_p)$ is the spin-s unitary representation of the pure boost $L_p$ that takes $k$ to $p$. Explicitly \cite{Polyzou},
\begin{equation}
L_p = \frac{1}{M} \begin{bmatrix}
p^0 && \mathbf{p}^T \\
\mathbf{p} && \delta_{ij}+\frac{p_i p_j}{M+p^0}
\end{bmatrix},    
\end{equation}
where $\mathbf{p}$ denotes the spatial part of $p$ and latin indices with values $1$, $2$, $3$ are used as spatial indices. 
The state $\ket{k, \sigma}$ is an eigenstate of both the momentum operator, $P$, and the total angular momentum in the $z$ direction, $J_z$
\begin{align}
P \ket{k, \sigma} =& k \ket{k, \sigma} \nonumber \\
J_z \ket{k, \sigma} =& \sigma \ket{k, \sigma}.     
\end{align}
It is important to note that the transformation that takes $k$ to $p$ is not unique. In fact, $L(p)R$, where $R$ is any three-dimensional rotation, has the same effect, since $R$ acts trivially on $k$. Different choices for $R$ lead to different definitions of momentum and spin states.  

We also wish to point out that, for arbitrary $p$, the state $\ket{p, \sigma}$ is no longer an eigenstate of $J_z$, so that $\sigma$ is {\it not} the label of the spin of the particle in the reference frame where it has momentum $p$ but, rather, in the reference frame where it is at rest. This remark is important in, for example, the context of spin measurements performed by a Stern-Gerlach apparatus. To illustrate this point suppose that we prepare a spin-$1/2$ particle in the state $\ket{k,\sigma = +z}$, and perform a quantum test with a Stern-Gerlach magnet oriented in the $z$-direction in the reference frame of the particle. The test consists in checking wether the particle has spin in the direction $+z$. In the rest frame of the particle it is absolutely certain that the particle will pass the test; however, if the particle has momentum $p$ in the reference frame of the magnet, there is a non-zero probability that it will be deflected in the $-z$ direction. Thus, a (normalised) state like
\begin{equation}
\label{MomentumSuperposition}
a\ket{p_1,+z}+b\ket{p_2,+z},
\end{equation}
cannot be interpreted as a spin eigenstate in this context, and taking a partial trace of the momentum degrees of freedom can lead to inconsistencies, as shown in \cite{SaldanhaVedralNJP}. This does not mean that the reduced spin density matrix obtained by tracing out the momentum degrees of freedom is useless for making physical predictions, as it gives the correct expectation values for suitably defined spin operators~\cite{Taillebois}. In our case, for example, the state (\ref{MomentumSuperposition}) is certain to pass a test corresponding to the projector $\ketbra{p_1,+z}{p_1,+z}+\ketbra{p_2,+z}{p_2,+z}$ and can therefore be understood as having spin $+z$ in this context. In this work we analyse spin-reduced density matrices formed by the partial trace method and write formally the Hilbert space of the particle as a tensor product of spin and momentum subspaces, $H = H_p\otimes H_s$, with the understanding that the reduced density matrix has to be interpreted in terms of adequate quantum tests.    

Consider now an observer whose reference frame is obtained by means of the Lorentz transformation $\Lambda^{-1}$ from the original reference frame. For this observer, the state $\ket{p, \sigma}$ is transformed under the spin-$s$ unitary representation of $\Lambda$, that is, $\ket{p, \sigma}\longrightarrow U(\Lambda)\ket{p, \sigma}$. We now find the explicit form of $U(\Lambda)$. From (\ref{WignerBasis}) and the group representation property it follows that
\begin{align}
U(\Lambda)\ket{p,\sigma} = & U(\Lambda)\,U(L_p) \ket{k, \sigma} \nonumber \\
= & U(L_{\Lambda_p})\,U(L^{-1}_{\Lambda_p}\Lambda L_p) \ket{k, \sigma} \nonumber \\
= & U(L_{\Lambda_p})\,U(W(\Lambda, \mathbf{p})) \ket{k, \sigma}.
\end{align}   
The transformation $W(\Lambda, \mathbf{p})$ is a pure rotation, since it leaves the rest frame four-momentum $k$ invariant: $W(\Lambda, \mathbf{p})\,k = k$. It is called the Wigner rotation corresponding to the Lorentz transformation $\Lambda$ and momentum $p$. In general, for any type of particle, Wigner rotations form a group, called the little group corresponding to momentum $k$. For the case of massive particles Wigner's little group is the rotation group $SO(3)$. As a consequence of the above equation, the momentum part of the state is transformed from $p$ to $\Lambda p$, and the spin part of the state changes under the action of $SO(3)$, according to
\begin{equation}
\label{Transformation}
U(\Lambda)\ket{p,\sigma} = \sum_{\sigma^\prime} D_{\sigma^\prime \sigma}(W(\Lambda, \mathbf{p})) \ket{\Lambda p, \sigma^\prime},   
\end{equation} 
where $D_{\sigma^\prime \sigma}(W(\Lambda, \mathbf{p}))$ is a spin-$s$ representation of the rotation $W(\Lambda, \mathbf{p})$. 

When considering two particle states it will be useful to look at transformation (\ref{Transformation}) with a different notation, separating the spin and the momentum parts of the system. We thus write  
\begin{equation}
U(\Lambda)\ket{p,\sigma} = \ket{\Lambda p}U_s(\Lambda,\mathbf{p}) \ket{\sigma},   
\end{equation}
where $U_s(\Lambda,\mathbf{p})\ket{\sigma} = \sum_{\sigma^\prime} D_{\sigma^\prime \sigma}(W(\Lambda, \mathbf{p})) \ket{\sigma^\prime}$. Considering again the state (\ref{MomentumSuperposition}) we see that it transforms as 
\begin{equation}
a\ket{p_1,+z}+b\ket{p_2,+z} \longrightarrow a\ket{\Lambda p_1}U_s(\Lambda,\mathbf{p}_1)\ket{+z} + b\ket{\Lambda p_2}U_s(\Lambda,\mathbf{p}_2)\ket{+z}. 
\end{equation}
Since we cannot write the final state as a tensor product of spin and momentum sectors we say that the Lorentz transformation has entangled the spin and the momentum. It is also said that Lorentz transformations do not preserve the tensor product structure of the Hilbert space. Of course, this spin-momentum entanglement is of a different nature as the usual particle-particle entanglement and should be understood in terms of concrete measurements. In our example, given by state (\ref{MomentumSuperposition}), we see that the transformed state is no longer an eigenstate of a projector of the form
\begin{equation}
\label{Projector}
\ketbra{\Lambda p_1,+n}{\Lambda p_1,+n}+\ketbra{\Lambda p_2,+n}{\Lambda p_2,+n}
\end{equation}
for any direction $n$, in contrast to state transformations given by pure rotations, where $U(\Lambda,\mathbf{p}_1)$ and $U(\Lambda,\mathbf{p}_2)$ are equal. Therefore, due to spin-momentum entanglement, the particle has a non-zero probability to fail a test for the operator (\ref{Projector}) for every possible value of $n$. This is reflected by the fact that the reduced spin density matrix is no longer a pure state when tracing out momenta. We want to make clear that the situation just described poses no problem for relativistic invariance: we are talking about two different experiments rather than a single experiment seen by two different inertial observers. 
 
\section{Two particles}
\label{three}
\noindent Consider now a pair of spin-$s$ distinguishable massive particles with momenta $p_1$ and $p_2$ according to the reference frame of some inertial observer. The state of the system in this reference frame is $\ket{p_1, \sigma_1}\otimes\ket{p_2, \sigma_2}$. This state is a basis element of the complete Hilbert space, which we decompose into two possible partitions 
\begin{align}
	H =& \,H_{A}\otimes H_{B} \nonumber\\
	=& \,(H_{A}\otimes H_{B})_{p}\otimes(H_{A}\otimes H_{B})_{s} \nonumber \\
	=& \, H_p \otimes H_s,
\end{align}
where $A$ and $B$ denote our two particles and $p$ and $s$ stand for the momentum and spin degrees of freedom, respectively. 

For the second inertial observer described in the previous section, the two particle system is described by 
\begin{align}
\label{Transformation2P}
U(\Lambda)\ket{p_1, \sigma_1}\otimes\ket{p_2, \sigma_2} =& U_1(\Lambda)\ket{p_1, \sigma_1}\otimes U_2(\Lambda)\ket{p_2, \sigma_2} \nonumber \\
=& \sum_{\sigma_1^\prime \sigma_2^\prime} D_{\sigma_1^\prime \sigma_1}(W(\Lambda, \mathbf{p}_1))D_{\sigma_2^\prime \sigma_2}(W(\Lambda, \mathbf{p}_2))\ket{p_1, \sigma_1^\prime}\otimes\ket{p_2, \sigma_2^\prime} \nonumber \\
=& \sum_{\sigma_1^\prime \sigma_2^\prime} \left(D(W(\Lambda, \mathbf{p}_1))\otimes D(W(\Lambda, \mathbf{p}_2))\right)_{\sigma_1^\prime \sigma_2^\prime, \, \sigma_1 \sigma_2} \ket{p_1, \sigma_1^\prime}\otimes\ket{p_2, \sigma_2^\prime}. 
\end{align}
The most important thing to note about this transformation is that it acts with a unitary operator in each particle subspace. By linearity, this will be true for an arbitrary initial state. As a consequence, entanglement between particles will always be conserved. This fact is fundamental for the consistency between quantum mechanics and special relativity as, for example, a violation of Bell's inequalities in one reference frame implies a violation of the inequalities in any other frame. In order for this to be true, entanglement between particles must be a relativistic invariant.
  
Having said that, we now analyse the transformation in the two-particle spin space. From eq.(\ref{Transformation2P}) we see that it is given by the tensor product of the representations of the Wigner rotations corresponding to each particle, i.e. we may rewrite eq.(\ref{Transformation2P}) as
\begin{equation}
U(\Lambda)\ket{p_1, p_2}\ket{\sigma_1, \sigma_2} = \ket{\Lambda p_1, \Lambda p_2}U_s(\Lambda, \mathbf{p}_1, \mathbf{p}_2)\ket{\sigma_1, \sigma_2},
\end{equation}
where $U_s(\Lambda, \mathbf{p}_1, \mathbf{p}_2) = D(W(\Lambda, \mathbf{p}_1))\otimes D(W(\Lambda, \mathbf{p}_2))$. From this result we can now find the group acting on the spin part of the system for a given pair of momenta $\mathbf{p}_1$ and $\mathbf{p}_2$. Since the transformation $D(W(\Lambda, \mathbf{p}_1))\otimes D(W(\Lambda, \mathbf{p}_2))$ depends on two $SO(3)$ elements, $W(\Lambda, \mathbf{p}_1)$ and $W(\Lambda, \mathbf{p}_2)$, it is a representation of the cartesian product $SO(3) \times SO(3)$. In general, for two groups $\mathcal{G}_1$ and $\mathcal{G}_2$ and two representations $D_1$ and $D_2$ acting on two vector spaces $V_1$ and $V_2$, that is, $D_1: \mathcal{G}_1 \longrightarrow GL(V_1)$ and $D_2: \mathcal{G}_2 \longrightarrow GL(V_2)$, we can construct the exterior tensor product representation 
\begin{equation}
D_1 \boxtimes D_2: \mathcal{G}_1 \times \mathcal{G}_2 \longrightarrow V_1 \otimes V_2,
\end{equation}
defined by 
\begin{equation}
(g_1, g_2) \mapsto D_1(g_1) \otimes D_2(g_2), 
\end{equation}
for all $g_1 \in \mathcal{G}_1$ and $g_2 \in \mathcal{G}_2$. In the case described above, $\mathcal{G}_1 = \mathcal{G}_2 = SO(3)$ and the representation acting on the spin space $H_s$ is an exterior product of the representations described in the last section.

We can ask for the invariant subspaces of these representations in order to find a natural division of the spin space for the physical situation described above. However, it is a known result of exterior tensor product representations that $D_1 \boxtimes D_2$ is irreducible if and only if $D_1$ and $D_2$ are irreducible. Since, by assumption, we have elementary particles, both representations of $SO(3)$ corresponding to $\mathbf{p}_1$ and $\mathbf{p}_2$ are irreducible. As a consequence the two-particle spin subspace has no nontrivial invariant subspaces. 

The situation is different, however, when both momenta are correlated. In the case where, say, $\mathbf{p}_2$ is a linear function of $\mathbf{p}_1$, i.e. $\mathbf{p}_2 = f_{lin}(\mathbf{p}_1)$, the representation of $SO(3)\times SO(3)$ becomes effectively a representation of $SO(3)$, since the elements $(W(\Lambda,\mathbf{p}),W(\Lambda, f_{lin}(\mathbf{p}))$ are in one to one correspondence with $W(\Lambda, \mathbf{p}) \in SO(3)$. In this case the representation will be in general reducible and there will be nontrivial subspaces of the two-particle spin space that transform amongst themselves under Lorentz boosts. 

In an EPR-like scenario, a pair of particles is created with $0$ total linear momentum, so that the functional relation between $\mathbf{p}_1$ and $\mathbf{p}_2$ is simply $\mathbf{p}_2 = -\mathbf{p}_1$. For this scenario, and a Lorentz boost $\Lambda$ in a given direction, the underlying group is actually $SO(2)$, since the Wigner rotation is along the same axis for both particles.

\section{Spin-momentum entanglement (Results)}
\label{four}

\subsection{General Results}
\noindent Let us closely analyse the situation described in the last paragraph of Section \ref{three}. We first state the physical scenario briefly, following previous treatments \cite{Alsing, Friis,Castro-RuizNahmad-Achar}, and then study the behaviour of spin-momentum entanglement in the light of the invariant subspaces that arise due to the correlation between momenta.

Let $p = (p^0, \mathbf{p})$ (and $-p$) be characterised by the rapidity $\mathbf{\eta}$, which is a three-vector that points in the direction of $\mathbf{p}$ and satisfies $M \sinh\vert \mathbf{\eta} \vert = \vert \mathbf{p} \vert$. Let $\Lambda$ be a pure boost perpendicular to the propagation direction and parametrised by the rapidity $\mathbf{\xi}$, so that $\tanh\vert \mathbf{\xi} \vert = \mathbf{v}$, with $v$ equal to the relative velocity between the reference frames. 

In this case the Wigner rotation $W(\Lambda,\mathbf{p})$ is along the axis defined by $\mathbf{\eta}\times\mathbf{\xi}$ and has a rotation angle $\Omega$, given by
\begin{equation}
\tan \Omega = \frac{\sinh\vert \mathbf{\eta} \vert \sinh\vert \mathbf{\xi} \vert}{\cosh\vert \mathbf{\eta} \vert + \cosh\vert \mathbf{\xi} \vert}.
\end{equation}    
The angle $\Omega$ is called the Wigner angle. The Wigner rotation corresponding to the opposite value of the momentum, $W(\Lambda,-\mathbf{p})$, is equal to the previous one but replacing $\Omega$ by $-\Omega$.
Then the rotation axes corresponding to $W(\Lambda,\mathbf{p})$ and $W(\Lambda, -\mathbf{p})$ are the same, and the underlying group that acts on the spin space is $SO(2)$. The group $SO(2)$ has one-dimensional irreducible representations of the form $e^{\mathrm{i} m \Omega}$. The representation induced in the two-particle spin-space by the Lorentz transformation must be reducible, since this space is $(2s+1)^2$-dimensional.

Based on this idea, we label the spin states in $H_s$ according to their transformation properties under representations of `rotations' of the form $W(\Lambda, \mathbf{p})\otimes W(\Lambda, -\mathbf{p})$. More precisely, we define the spin state $\ket{m,\alpha}$ as an element which carries the $m$ representation of $SO(2)$, i.e.,
\begin{equation}
\label{InvariantBasis}
U(\Lambda)\ket{p,-p}\ket{m,\alpha} = e^{\mathrm{i}m\Omega}\ket{\Lambda p, -\Lambda p}\ket{m, \alpha}.    
\end{equation}
The label $\alpha$ is used in order to take into account the different times that the same irreducible representation of $SO(2)$, labeled by $m$, appears in the spin space. The number of different $\alpha$'s for a given $m$, that is, the multiplicity of the representation $m$, is calculated in \cite{Castro-RuizNahmad-Achar} and shown to be $a_m = 2 s +1 - \vert m \vert$. Therefore the transformation $W(\Lambda, \mathbf{p})\otimes W(\Lambda, -\mathbf{p})$ is diagonal in the $\{ \ket{m, \alpha} \}$ basis. Moreover, since the transformation that diagonalises $W(\Lambda, \mathbf{p})\otimes W(\Lambda, -\mathbf{p})$ is unitary, $\{ \ket{m, \alpha} \}$ is an orthonormal set,
\begin{equation}
\braket{m, \alpha}{m^\prime, \alpha^\prime} = \delta_{m m^\prime}\delta_{\alpha \alpha^\prime}.    
\end{equation} 

We now quantify how the Lorentz boost $\Lambda$ entangles the spin and the momentum of the most general spin $s$ initial spin-state. Since the entanglement change is $0$ for initial momentum states that are not in a superposition \cite{Friis}, we take the initial momentum state to be in the homogeneous superposition $\left( \ket{p,-p}+ \ket{-p,p} \right)/\sqrt{2}$. Our total initial state is then
\begin{equation}
\label{InitialState}
\ket{\psi_i} = \frac{\ket{p,-p} + \ket{-p,p}}{\sqrt{2}} \sum_{m = -2s}^{2s} \sum_{\alpha = 1}^{a_m} c_{m \alpha}\ket{m,\alpha}.  
\end{equation}
Using equation (\ref{InvariantBasis}) in equation (\ref{InitialState}) we find the final state $\ket{\psi_f} = U(\Lambda)\ket{\psi_i}$ to be
\begin{align}
\ket{\psi_f} = \sum_{m = -2s}^{2s} \frac{ e^{\mathrm{i}m\Omega} \ket{\Lambda p,- \Lambda p} + e^{-\mathrm{i}m\Omega} \ket{-\Lambda p,\Lambda p} }{\sqrt{2}} \sum_{\alpha = 1}^{a_m} c_{m\alpha} \ket{m,\alpha}.   
\label{FinalState}   
\end{align}
From expression (\ref{FinalState}) we can see immediately that, for the case of a single value of $m$, i.e. for coefficients $c_{m,\alpha} = \delta_{m m_0}c_\alpha$, the spin sector of the system factors out and remains unchanged. There is therefore no entanglement between spin and momentum and, as a consequence, the entanglement between single-particle spins remains invariant as well, a fact that might be important for future applications of quantum information protocols in relativistic scenarios.  

Let us now calculate, as a measure of the entanglement between momentum and spin, the linear entropy with respect to the momentum-spin partition of the Hilbert space \cite{Friis}
\begin{equation}
\label{LinearEntropy}
E=\sum_{i} \left(1-Tr\left(\rho_{i}^2\right)\right),    
\end{equation}   
where $\rho_i$ is obtained by tracing out the momentum or spin degrees of freedom from the total density matrix $\rho = \ketbra{\psi_f}{\psi_f}$. To simplify calculations we consider sharp momentum distributions approximated by plane-wave states so that, effectively, $\braket{p_i}{p_j}=\delta_{ij}$. 
Using equations (\ref{FinalState}) and (\ref{LinearEntropy}) we find that the linear entropy after the Lorentz boost is given by
\begin{equation}
\label{Entanglement}
E = 2\left(1-\sum_{m = -2s}^{2s}\sum_{m^\prime = -2s}^{2s} \,\sum_{\alpha = 1}^{a_m} \,\,\sum_{\alpha^\prime = 1}^{a_m^\prime} \vert c_{m \alpha}\vert^2 \vert c_{m^\prime \alpha^\prime}\vert^2 \cos^2(m-m^\prime)\Omega\right). 
\end{equation} 

\subsection{Examples: two parametrisations}
In order to analyse the behaviour of several spin sates at a time, several parametrisations of initial spin states were proposed in references \cite{Friis} and \cite{Castro-RuizNahmad-Achar}. The parametrisations were defined such that for every definite value of the parameters there was certain initial spin state. In this way, entanglement was calculated as a function of the parameters. Nevertheless, none of the spin state parametrisations proposed was `natural', in the sense that different parametrisations do not mix under Lorentz boosts. For example, it could seem natural to choose to parametrise the spin states according the the (non relativistic) total spin quantum number $s$ (corresponding to the operator $S^2 = S_x^2+S_y^2+S_z^2$), so that states with $s = 0$ belong to one parametrisation, states with $s = 1$ to another, and so on. However, the label $s$ is of course not conserved when applying Lorentz boosts on the states and different parametrisations mix. Now, according to the invariant basis states of equation (\ref{InvariantBasis}), all the spin states that transform under the same $SO(2)$ representation stay invariant when acting on them with a Lorentz transformation. Therefore the `natural' set of states to choose are the invariant subspaces labeled by different values of $m$. For these sets of states the question of how spin and momentum get entangled after the Lorentz boost is trivial: entanglement is zero for all linear combinations of the form $\sum_\alpha c_\alpha \ket{m, \alpha}$. It follows that all the information about how momentum and spin get entangled lies in the differences $m-m^\prime$ of the representation labels, as equation (\ref{Entanglement}) shows. 

\begin{figure}[h]
\centering
\scalebox{0.37}{\includegraphics{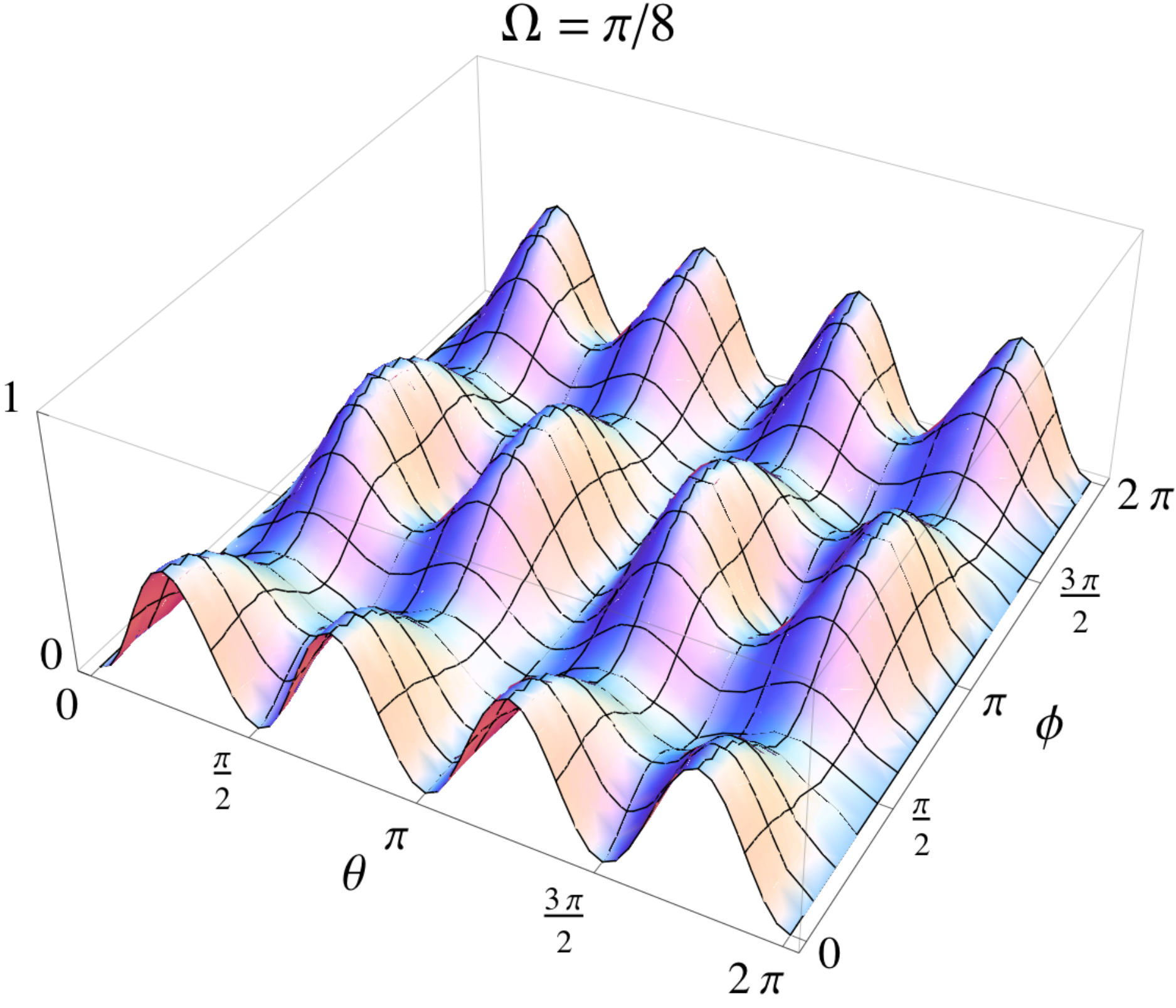}} \quad
\scalebox{0.37}{\includegraphics{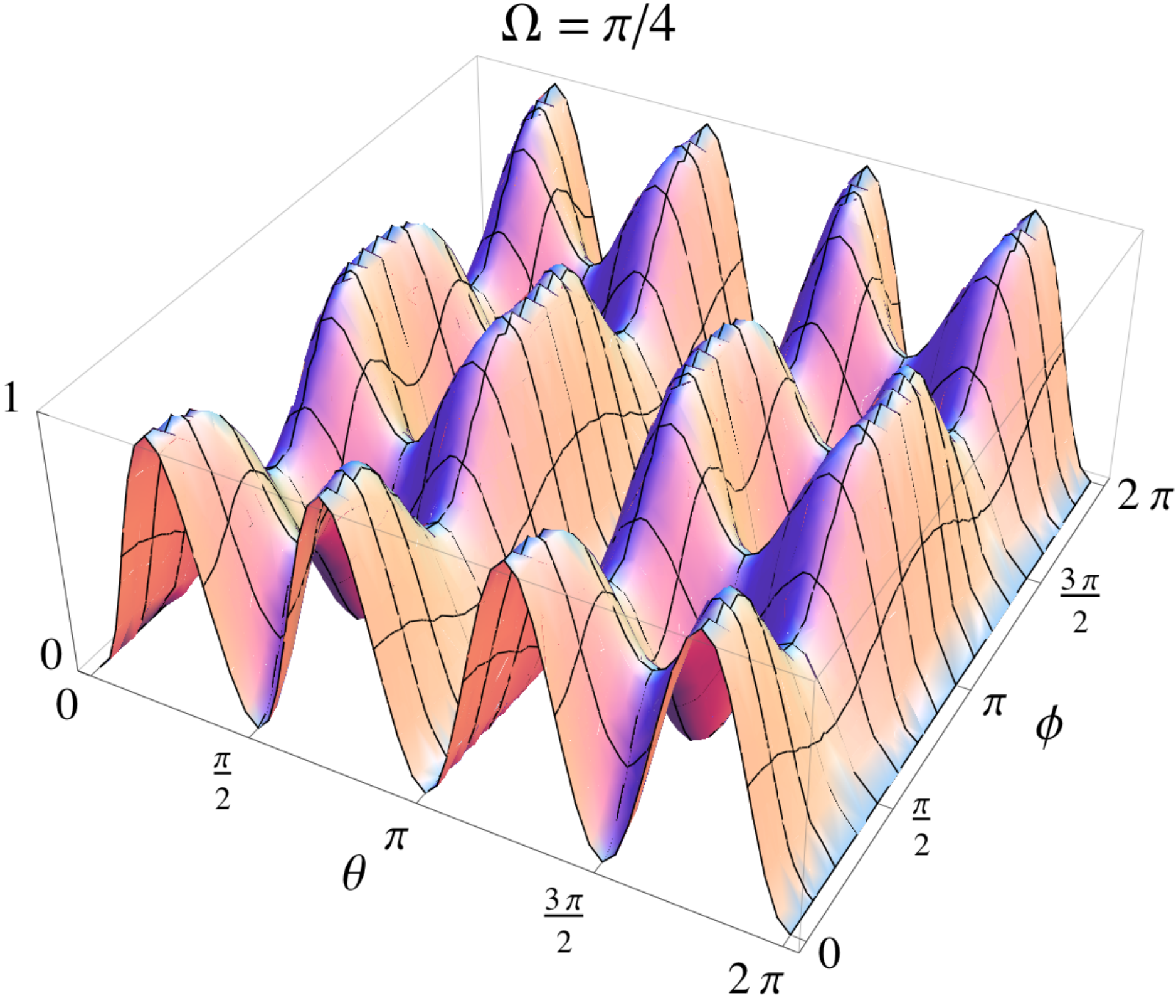}} \\
\scalebox{0.37}{\includegraphics{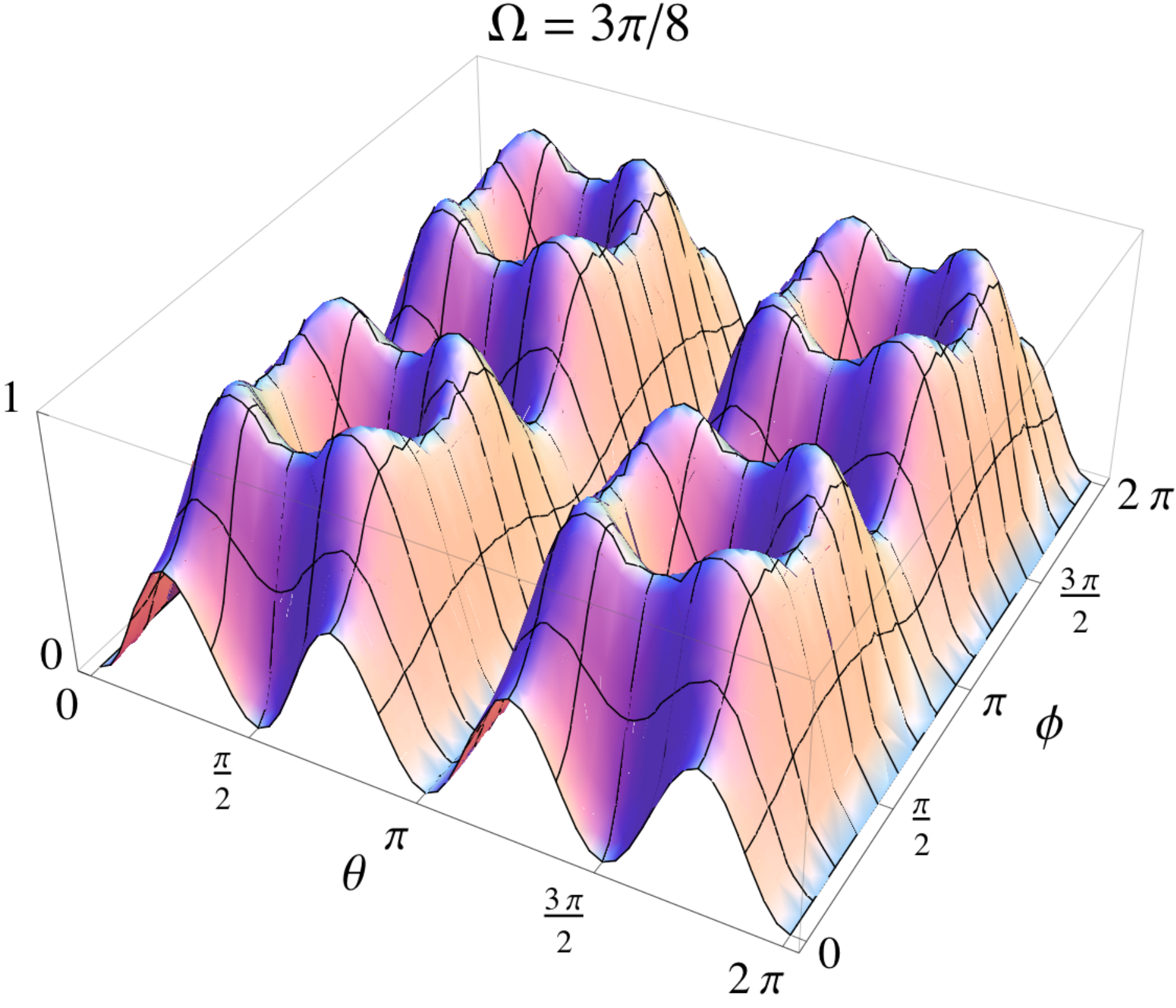}} \quad
\scalebox{0.37}{\includegraphics{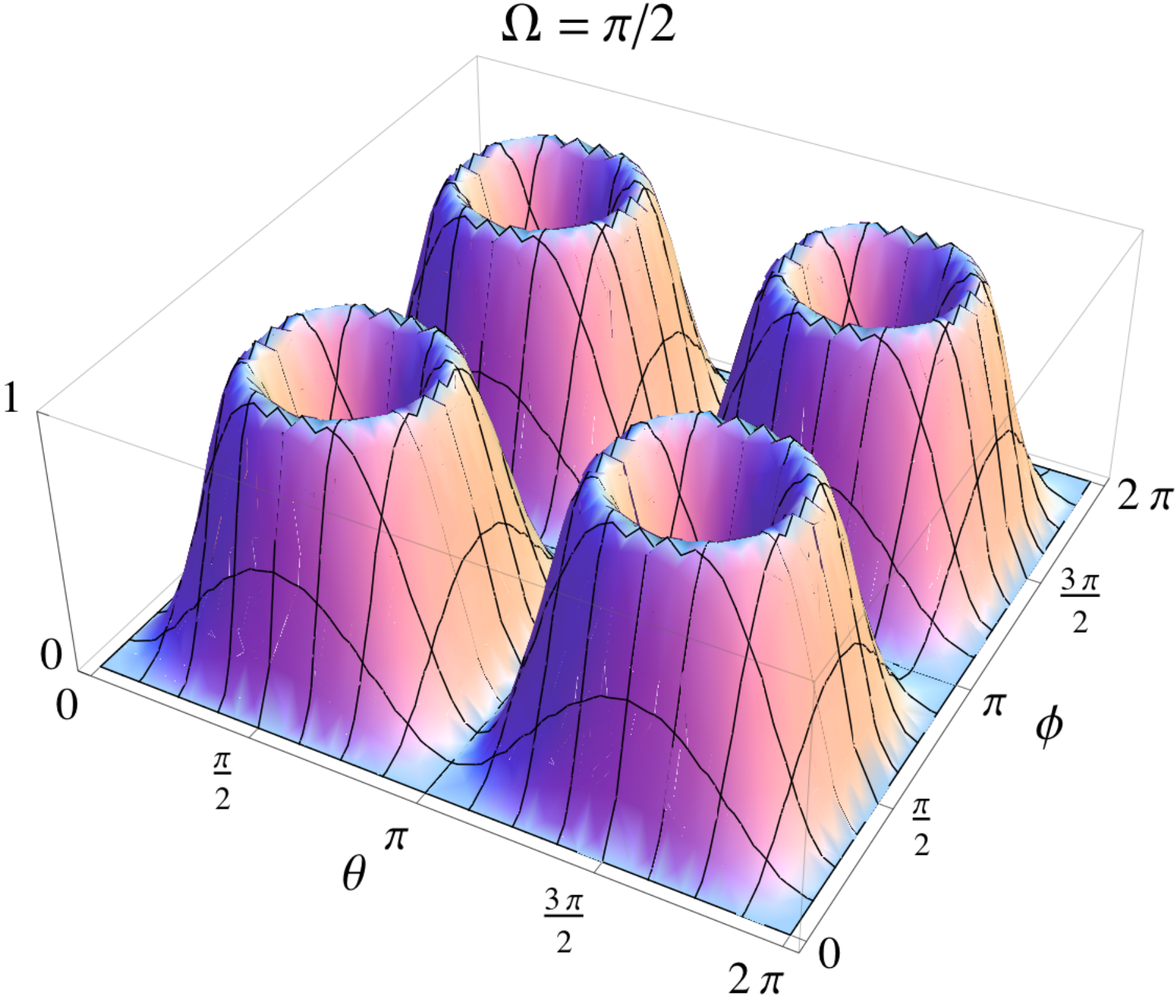}}
\caption{Spin-momentum entanglement for the states parametrised by eq. (\ref{Eq_ParamA}). }
\label{ParamA}
\end{figure}

We now illustrate the spin-momentum entanglement for superpositions of different values of $m$. Figure \ref{ParamA} shows the entanglement after the Lorentz boost for the states given by the parametrisation 
\begin{equation}
\label{Eq_ParamA}
\ket{\psi_s} = \sin\theta \cos\phi \ket{m = 1}+ \sin\theta \sin\phi \ket{m = 0} + \cos\theta \ket{m = -1},  
\end{equation} 
where we have ignored the value of $\alpha$ since it plays no role in entanglement.   

Note that, so far, no reference has been made to the total spin of the particles. In fact, the representations $m = 1, \ 0, \  -1$ are present for every $s>1/2$, so that Figure~\ref{ParamA} can describe particles of arbitrary spin. For the case $s = 1/2$, the invariant states are given explicitly by
\begin{subequations} 
\begin{align}
\ket{\psi^+} && U_s(\Lambda, \mathbf{p}, -\mathbf{p})\ket{\psi^+} =& \ket{\psi^+}  \\
\ket{\phi^-} && U_s(\Lambda, \mathbf{p}, -\mathbf{p})\ket{\phi^-} =& \ket{\phi^-}  \\
\ket{\chi^+} && U_s(\Lambda, \mathbf{p}, -\mathbf{p})\ket{\chi^+} =& e^{\mathrm{i}\Omega}\ket{\chi^+} \\
\ket{\chi^-} && U_s(\Lambda, \mathbf{p}, -\mathbf{p})\ket{\chi^-} =& e^{-\mathrm{i}\Omega}\ket{\chi^-},    
\end{align} 
\end{subequations}  
where
\begin{subequations}
\begin{align}
\ket{\psi^+} =& \frac{1}{\sqrt{2}}(\ket{+z,-z}+\ket{-z,+z}) \\
\ket{\psi^-} =& \frac{1}{\sqrt{2}}(\ket{+z,-z}-\ket{-z,+z}) \\
\ket{\phi^+} =& \frac{1}{\sqrt{2}}(\ket{+z,+z}+\ket{-z,-z}) \\
\ket{\phi^-} =& \frac{1}{\sqrt{2}}(\ket{+z,+z}-\ket{-z,-z})
\end{align}
\end{subequations}
are the well-known Bell states, and
\begin{subequations}
\begin{align}
\ket{\chi^+} =& \frac{1}{\sqrt{2}}(\ket{\phi^+}+\mathrm{i}\ket{\psi^-}) \\
\ket{\chi^-} =& \frac{1}{\sqrt{2}}(\ket{\phi^+}-\mathrm{i}\ket{\psi^-}).
\end{align}
\end{subequations}

The figure shows the spin-momentum entanglement for increasing Wigner angles. Entanglement is $0$ for vanishing $\Omega$ and increases gradually as $\Omega$ grows, as can be seen in the case $\Omega = \pi/8$. Note how invariant states $\ket{m=1}$ ($\theta = \pi/2$, $\phi = 0$), $\ket{m=0}$ ($\theta = \pi/2$, $\phi = \pi/2$) and $\ket{m=-1}$ ($\theta = 0$) always have $0$ entanglement. 

For thse spin-$1/2$ case, analysed in \cite{Friis}, the state $\ket{m=0}$ corresponds to either $\ket{\psi^+}$ or $\ket{\phi^-}$, which are maximally entangled states. These spin states remain exactly the same before an after the Lorentz boost and therefore are ideal candidates for transmitting quantum information in a situation where we want both observers to describe the same spin state, regardless the rapidity of the particles $\vert \mathbf{\eta} \vert$ or the strength of the boost $\vert \mathbf{\chi} \vert$. On the other hand, for situations where we want the state to change and therefore need the spin and momentum to get entangled, we note that the states for which entanglement is grater, that is, those corresponding to maxima in Figure~\ref{ParamA}, depend strongly on the Wigner angle. For $\Omega = \pi/8$ and $\pi/4$ there are maxima corresponding to the states $\ket{\phi^+} = (\ket{\chi^+}+\ket{\chi^-})/\sqrt{2}$ (with $\theta = \pi/4$, $\phi = 0$), and $\ket{\psi^-} = (\ket{\chi^+}-\ket{\chi^-})/\sqrt{2}$ (with $\theta = 3\pi/4$, $\phi = 0$), while for $\Omega = \pi/2$, corresponding to the limit of the speed of light, both of these states have $0$ entanglement. 

For the spin-$1$ case there are more options to choose from as representatives for the different $m$-representations. For example, the states 
\begin{subequations}
\begin{align}
\ket{\psi_1} =& \frac{1}{\sqrt{3}}\left( \ket{1,1} -\ket{0,0} +\ket{-1,-1} \right) \label{Entangled1}\\
\ket{\psi_2} =& \frac{1}{\sqrt{3}}\left( \ket{1,-1} +\ket{0,0} +\ket{-1,1} \right) \label{Entangled2}\\ 
\ket{\psi_3} =& \frac{1}{2}\left( \ket{1} +\ket{-1} \right)\left( \ket{1} +\ket{-1} \right) \label{Separable},   
\end{align}    
\end{subequations}
where $1$, $0$ and $-1$ denote the spin projection along the $z$-axis, carry the representation $m=0$ of $SO(2)$. Note that (\ref{Entangled1}) and (\ref{Entangled2}) are maximally entangled spin-$1$ states, while (\ref{Separable}) is separable. 

For the case $m = \pm 1$ and $s =1$ we have, for example, the states
\begin{subequations}
\begin{align}
\ket{\beta_1} =& \frac{1}{\sqrt{2}}\left(\ket{1,-1}-\ket{-1,1}\right) \\ 
\ket{\beta_2} =& \frac{1}{2}\left(\ket{0}\left(\ket{1}-\ket{-1}\right)+\left(\ket{1}-\ket{-1}\right)\ket{0}\right),   
\end{align}
\end{subequations}
that transform amongst themselves
\begin{equation}
U_s(\Lambda,\mathbf{p},-\mathbf{p})\begin{bmatrix} 
\ket{\beta_1} \\
\ket{\beta_2}
\end{bmatrix} 
= 
\begin{bmatrix} 
\cos\Omega & \sin\Omega \\
-\sin\Omega & \cos\Omega
\end{bmatrix} 
\begin{bmatrix} 
\ket{\beta_1} \\
\ket{\beta_2}
\end{bmatrix}. 
\end{equation}
As a consequence, the linear combinations $(\ket{\beta_1}+\mathrm{i}\ket{\beta_2})$ and $(\ket{\beta_1}-\mathrm{i}\ket{\beta_2})$ carry the representations $m = 1$ and $m = -1$, respectively. From these examples we note that, while the analysis of spin-momentum entanglement is independent of the spin of the particles and of the particular realisation of the different states that carry $SO(2)$ representations, such realisations have to be taken into account when studying entanglement between pairs of spins, since the characteristic of the states (maximally entangled, partially entangled, separable) may differ strongly from case to case. Nevertheless, we can say safely that if the spin state remains unchanged, as is the case for superpositions that transform under the same representation, then the entanglement between spins will also remain unchanged, no matter its value.  

\begin{figure}[h]
\centering
\scalebox{0.37}{\includegraphics{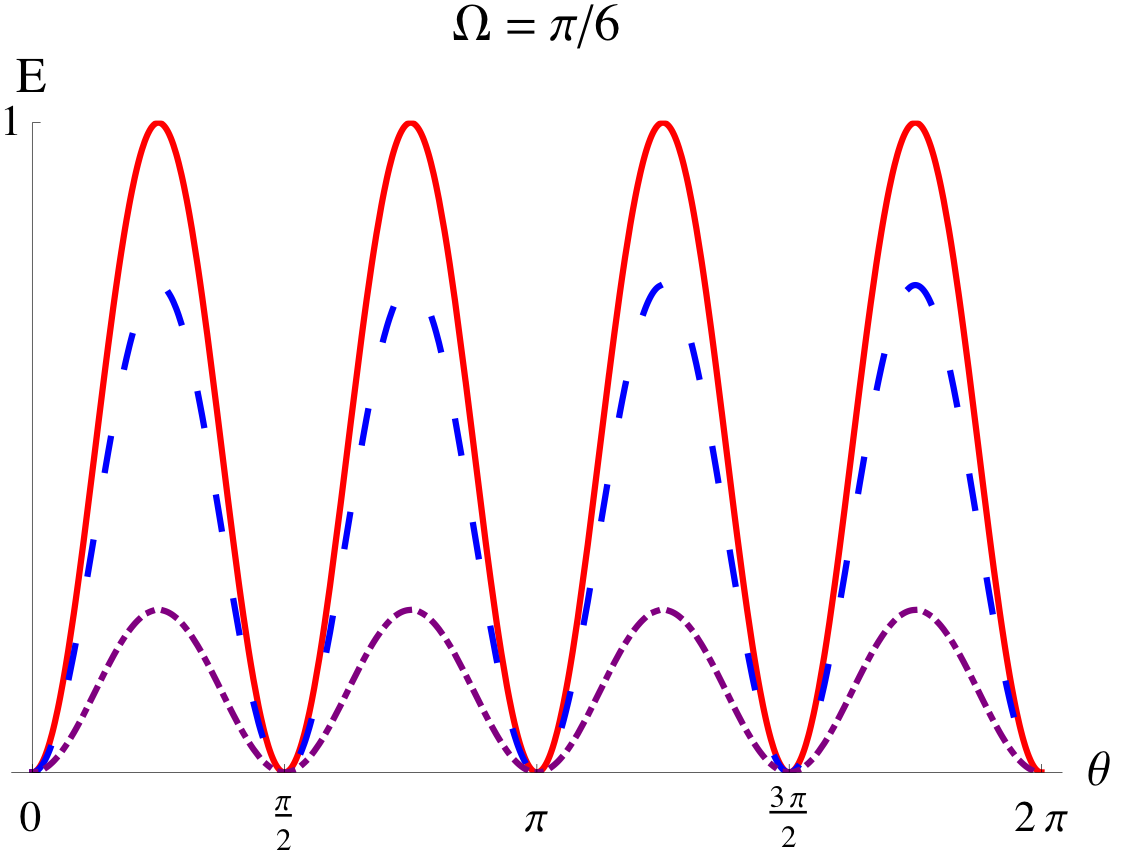}} \quad
\scalebox{0.37}{\includegraphics{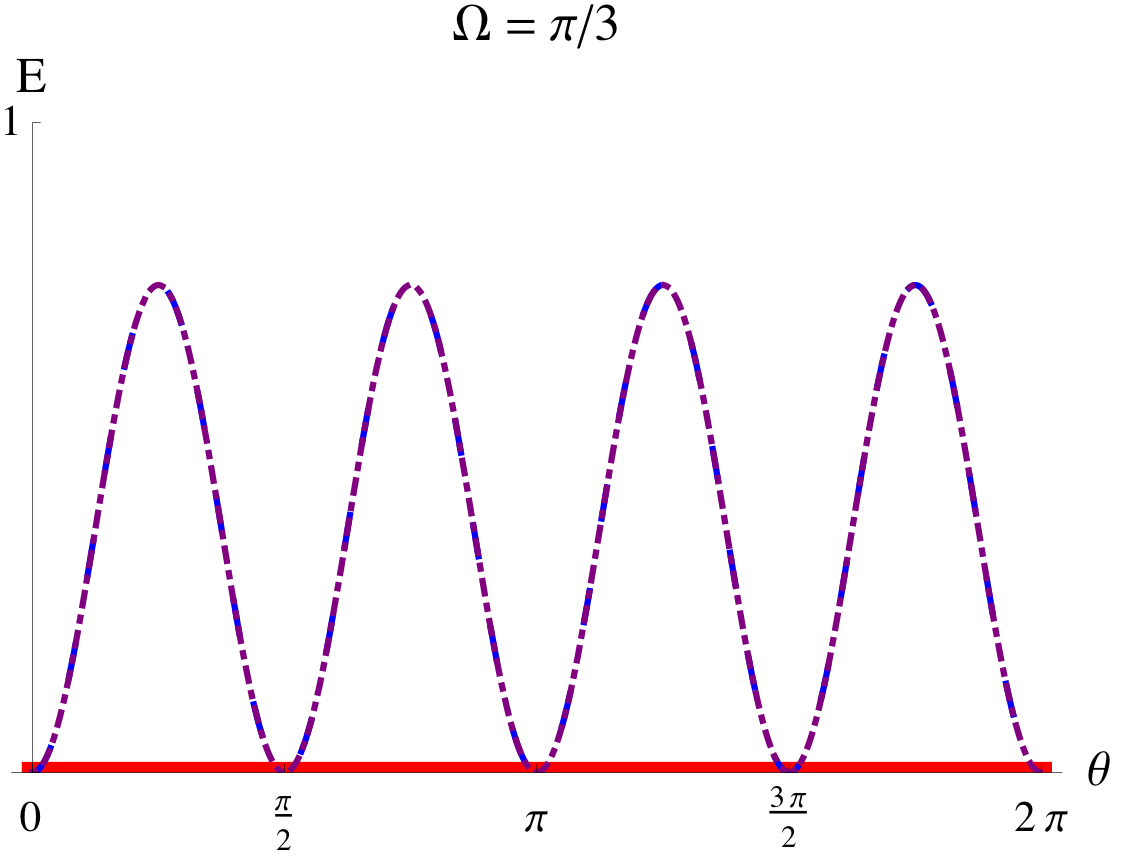}} \quad 
\scalebox{0.37}{\includegraphics{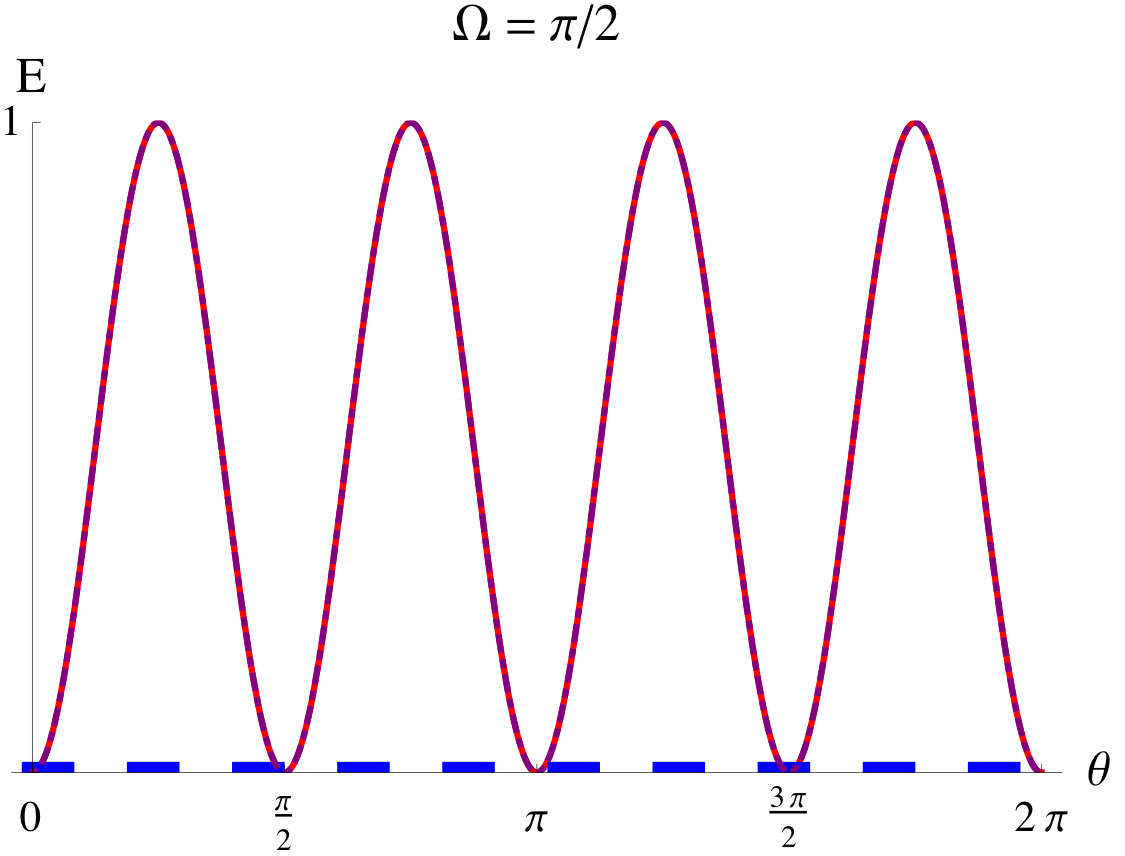}} 
\caption{Spin-momentum entanglement for the superposition states in (\ref{Eq_ParamB}). \textbf{Solid line} (red online): $m-n = 3$. \textbf{Dashed line} (blue online) $m-n = 4$. \textbf{Dot-dashed line} (purple online): $m-n = 5$. See text for details.}
\label{ParamB}
\end{figure}

We now analyse the entanglement for a general superposition of two different representations, labeled by $m$ and $n$, without any reference to the spin of the particles. Figure \ref{ParamB} shows spin momentum entanglement for the general superposition

\begin{equation}
\label{Eq_ParamB}
\cos\theta \ket{m}+ e^{\mathrm{i}\phi} \sin\theta \ket{n},  
\end{equation}  
where we have again ignored the label $\alpha$. We choose the three different values of $3$, $4$ and $5$ for $m-n$ since the cases $m-n = 0, \ 1, $ and $2$ are already illustrated as particular cases in Figure~\ref{ParamA}.

The first thing to note directly from equation (\ref{Entanglement}) is that, since entanglement depends only on the squared amplitudes of the state, the relative phase $\phi$ is irrelevant and spin-momentum entanglement is only a function of $\theta$. Explicitly, entanglement takes the simple form
\begin{equation}
E = \sin^2 2\theta \sin^2(m-n)\Omega  
\end{equation}

Again, as in the case of equation (\ref{Eq_ParamA}), entanglement is invariant for $\ket{m}$ ($\theta =0$) and for $\ket{n}$ ($\theta =\pi/2$). As we can see in Figure~\ref{ParamB}, maxima are always at $\theta = n\pi/4$, with $n$ an integer, for all values of $m-n$. This corresponds to the states 
\begin{equation}
\label{MaximumEntanglement}
\frac{1}{\sqrt{2}}(\ket{m}+e^{\mathrm{i}\phi}\ket{n}).
\end{equation} 
Spin states of this form are the ones that exhibit the maximum spin-momentum entanglement and therefore can be used for situations where the sender wants to transmit information encoded in the spin degrees of freedom to a particular receiver, who has a definite relative velocity with respect to his/her reference frame. Since entanglement depends on the boost velocity, the sender can prepare the state in such a way that the desired amount of entanglement, or the desired boosted spin state, is achieved only for the particular velocity of the receiver.
Moreover, entanglement also depends on $m-n$. For a relatively weak boost, $\Omega = \pi/6$, the states with $m-n=3$ have the maximal amount of entanglement (top left of Figure~\ref{ParamB}), while for $\Omega = \pi/3$ these states have $0$ entanglement for all values of $\theta$ (top right of Figure~\ref{ParamB}). For this last value of the Wigner angle, the cases $m-n =4$ and $m-n = 5$ are equivalent. When the Wigner angle corresponds to the limit of the speed of light, $\Omega = \pi/2$ (bottom of figure \ref{ParamB}), the state with $m-n = 4$ has no spin-momentum entanglement, while the cases $m-n = 3$ and $m-n = 5$ behave in the same way and have maximal entanglement for states of the form given by eq. (\ref{MaximumEntanglement}).

\section{Conclusions}
\label{five}
We have analysed the transformation properties of two-particle systems under Lorentz transformations from a quantum information perspective. We focused on the transformation corresponding to the spin degrees of freedom and showed that, in general, the spin subspace carries an exterior tensor product representation of $SO(3)\times SO(3)$. For arbitrary momenta this representation is irreducible but, interestingly, the representation becomes reducible for correlated momenta since the underlying group that acts on the spin space in this case becomes $SO(3)$. The states that span irreducible subspaces of $SO(3)$ have interesting properties since they transform amongst themselves under a Lorentz boost and are therefore good candidates for encoding quantum information in relativistic settings. For an EPR-like case, where the momenta of the particles are equal and opposite, the situation simplifies even more since the group that acts on the spin space is $SO(2)$, which has one-dimensional representations of the form $e^{\mathrm{i}m\Omega}$. We have analysed the transformation induced by the Lorentz boost in the spin space using a basis formed by states that carry representations of $SO(2)$ labeled by $m$. The transformation of the spin states and therefore their entanglement properties are independent of the total spin of the particle and a general treatment in terms of invariant states is possible. Superpositions of spin states that transform according to the same value of $m$ remain unchanged after the boost and therefore the initial entanglement between individual spins is invariant.

The problem with encoding information into the spin and momentum degrees of freedom is that, since they become entangled as seen by different relativistic observers, the decoding of the said information is not trivial (or perhaps even possible). However, linear superpositions of states that carry the same representation of $SO(2)$ remain invariant under Lorentz boosts, thus offering the opportunity to encode/decode information regardless of the observer. On the other hand, one may wish to encode information into a state that only a particular observer will be able to decode, and the state may then be prepared so that the particular observer with its proper velocity receives the state with the wanted measure of entanglement in order to decode it; in this case superpositions of states that carry different representations of $SO(2)$ are appropriate.

As the entanglement between spin and momentum is most naturally analysed in terms of superpositions of states with different values of $m$, the basis of states used in this work to study transformations of spin states under Lorentz boosts is a good candidate for building quantum communication protocols in relativistic scenarios.

\section*{Acknowledgements}
This work was partially supported by DGAPA-UNAM under project IN101614.

\end{document}